\documentclass[journal,onecolumn]{IEEEtran}
\usepackage{setspace} 
\ifCLASSINFOpdf
  
\else
  
\fi

\usepackage{subcaption}
\usepackage{float}
\usepackage{tabularx}
\usepackage{mathtools}
\usepackage{empheq}
\usepackage{multirow}
\usepackage{verbatim}
\usepackage[ruled,linesnumbered,noend]{algorithm2e}
\usepackage{tikz}
\usepackage{microtype}
\usepackage[capitalise,noabbrev]{cleveref}
\usepackage{xcolor} 
\usepackage{colortbl}
\usepackage{booktabs}
\newcommand{\beqa}{\begin{eqnarray}}
	\newcommand{\eeqa}{\end{eqnarray}}

\begin{document}

\captionsetup{font=normalsize}
\include{defn}

\title{Cybersecurity Threats to Power Grid Operations from the Demand-Side Response Ecosystem}

\author{Subhash~Lakshminarayana~\IEEEmembership{Senior Member, IEEE}, Yexiang Chen~\IEEEmembership{Member, IEEE}, Carsten Maple, Andrew Larkins, Daryl Flack, Christopher Few, Kenny-Awuson David, and Anurag. K. Srivastava~\IEEEmembership{Fellow, IEEE} 
\thanks{ Corresponding author: Subhash~Lakshminarayana, subhash.lakshminarayana@warwick.ac.uk, School of Engineering, University of Warwick, Coventry - CV47AL, UK. \\
This work was supported by the PETRAS National Centre of Excellence for IoT Systems Cybersecurity through the U.K. EPSRC
under Grant EP/S035362/1.}
}

\maketitle

\begin{abstract}
This article focuses on cyber security threats from IoT-enabled energy smart appliances (ESAs) such as smart heat pumps, electric vehicle chargers, etc., to power grid operations. It presents an in-depth analysis of the demand side threats, including (i) an overview of the vulnerabilities in ESAs and the wider risk from the demand-side response (DSR) ecosystem, (ii) key factors influencing the attack impact on power grid operations, (iii) measures to improve the cyber-physical resilience of power grids, putting them in the context of ongoing efforts from the industry and regulatory bodies worldwide.
\end{abstract}

\section{Introduction}
\label{sec:Intro}
Demand side response (DSR) is seen as a critical element in achieving the net-zero goals set by several nations worldwide. It refers to the ability to shift load away from the peak demand periods and potentially align them with renewable energy generation. Energy smart appliances (ESAs), such as smart heat pumps, electric vehicle charging stations, etc., can offer DSR. The connectivity features of these smart devices can be leveraged to control them remotely. This can enable scheduling their usage during off-peak periods or provide on-demand services (e.g., increase/decrease demand by direct load control) while respecting the consumer’s usage preferences and giving them “override” privileges whenever needed. 

\begin{figure}[t]
    \centering
    \includegraphics[width=0.8\linewidth]{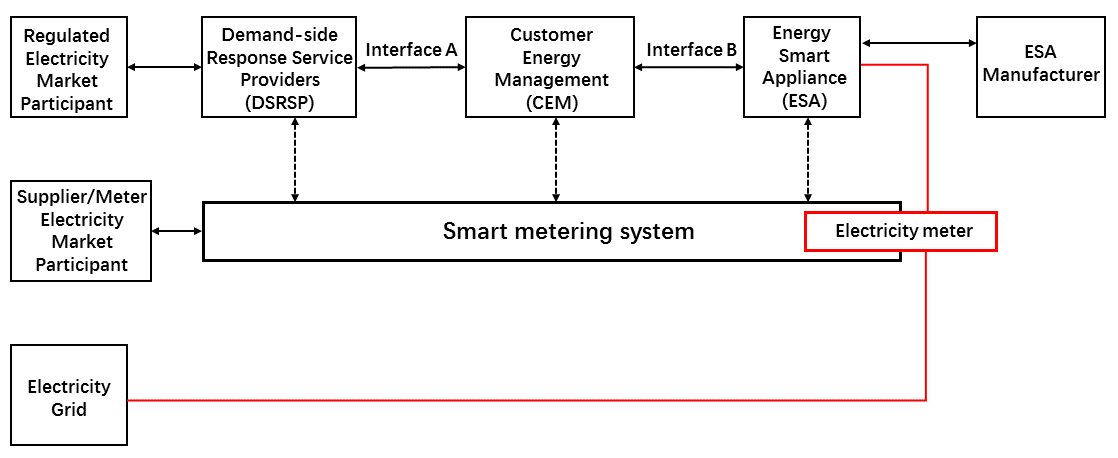}
    \caption{Logical DSR architecture and communications connections. (Figure adapted from British Standards Institution)}
    \label{fig: DSR_Log}
    \vspace{1cm}
\end{figure}

Fig. ~\ref{fig: DSR_Log} shows logical DSR architecture and the entities involved in DSR. Intermediary organisations called “DSR service providers” are responsible for providing DSR by directly (e.g., direct load control) or indirectly (e.g., through pricing schemes) controlling ESAs. Customer energy management (CEM) is a logical entity used to manage ESAs in order to provide DSR. The entities involved in the DSR ecosystem are connected via communication interfaces (i.e., Interface A and B in Fig. ~\ref{fig: DSR_Log}). Finally, the ESA is also connected to the manufacturer for management purposes, such as monitoring, firmware updates, etc.

\begin{figure}[t]
    \centering
    \includegraphics[width=0.8\linewidth]{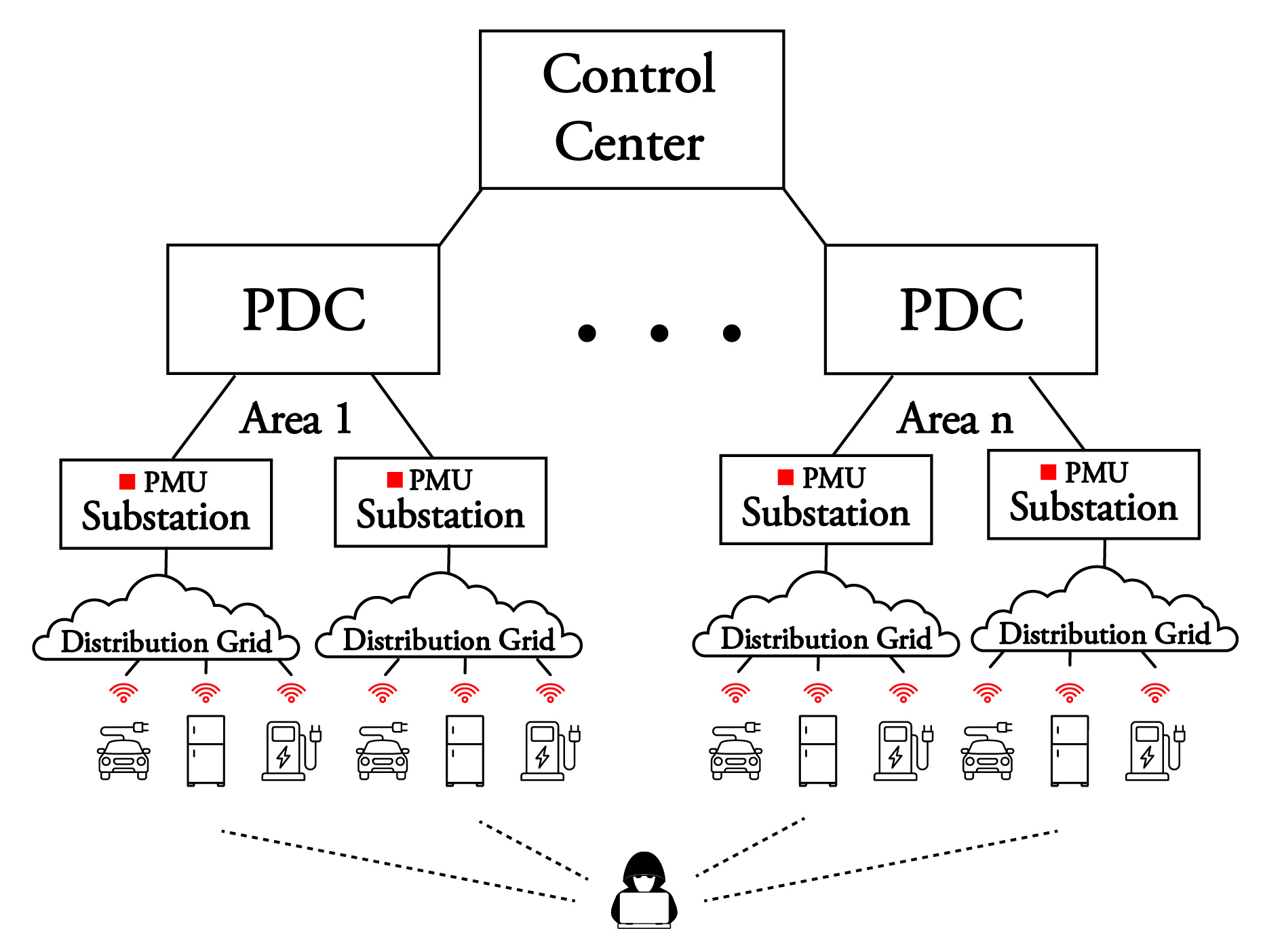}
    \caption{Illustration of demand-side cyber threats against power grid operations. The figure also shows a monitoring framework at bulk power grid level. PMU - Phasor measurement unit, PDC - Phasor data concentrator. }
    \label{fig:PMU_Hierarchy2}
    \vspace{1cm}
\end{figure}

While the benefits of ESAs are undeniable, they also pose cyber security risks. Several security vulnerabilities have been discovered in ESAs and the associated DSR communication interfaces ({weak authentication mechanisms, lack of security updates, etc.,} see Section~\ref{sec:TARA} for more details). These vulnerabilities can pose a serious threat to consumer’s privacy and security. More importantly, as the ESA penetration levels increase, they may become a new attack surface to target power grids. 

The balance between the supply (i.e., generation) and the demand (i.e., loads) is critical for power grid operations. The specific focus of this article is cyber threats that can lead to a sudden surge or drop in demand, causing disruptions to power grid operations. This can happen due to (i) load-altering attacks (LAAs) that compromise ESAs or the ESA communication interfaces and change the load control settings of a large number of ESAs in a coordinated manner, (ii) consumers acting in a coordinated but unexpected way, e.g., in response to an external prompt, and or (iii) ESAs responding at scale in a coordinated but unexpected way, e.g., due to misconfiguration, coding errors or an external signal. A pictorial depiction of LAAs is presented in Figure~\ref{fig:PMU_Hierarchy2}.

Real-world and isolated instances of cyber-attacks compromising ESAs have already been observed, such as the compromise of the EV charger display to show inappropriate images \cite{EVHack}. To the best of our knowledge, cyber-attacks leading to large-scale changes in the power demand (amounting to several mega-watts) have not been witnessed. However, sudden surges due to the synchronized actions from several customers are frequently observed during the “TV pick-up” effect in the GB grid. For instance, during England’s victory over Germany in UEFA EURO 2020, National Grid witnessed a “1 GW pick-up in electricity demand at half-time in the match, and around a 1.6~GW pick-up after full-time (equivalent power to around 320 million light bulbs and 888,000 kettles)’’. The “Pan-India Lights Switch Off Event” on 5 April 2020, in which the Indian government urged citizens to switch off electric lights and light lamps to show their support in the fight against the
COVID pandemic resulted in a nationwide demand drop of 31 GW.
As these were anticipated events (such as the TV pickup effect), the forecasting teams working with the grid operators were able to predict the demand surge/drop and suitable balancing mechanisms, such as pumped storage hydroelectric power stations, were dispatched to maintain the system balance. However, in case of a cyber event resulting in large-scale load changes, such surges/drops in demand may be completely unanticipated, potentially causing significant disruption.

In power grid security literature, the majority of existing works focus on utility-side cyber attacks and the associated SCADA system security \cite{MashimeNOW2023}. These attacks are well understood, thanks to the several years of research efforts on this topic. Security guidelines to mitigate these risks have been proposed as standards and regulations (e.g., NERC-CIP regulations, IEC 62351 standards, etc.). In contrast, the threats posed by demand-side appliances on power grid operations have received little attention. While the local impact of such exploits, such as threats to user privacy and/or increase in household electricity bills, etc., has been studied in the past \cite{AsgharSMPrivacy2017}, the global impact (on power grid operations) of simultaneously controlling a large number of IoT devices has only recently been studied in the research literature. Existing works \cite{Soltan2018, CardenasHahnIoT2020, LakshIoT2021, SingerSP2023, JahangirIoTJ2023} focus on individual technical aspects of LAAs. \cite{ayub2023secure} highlights vulnerabilities in smart grids due to IoT integration and proposes a blockchain-based authentication method for secure demand response management. \cite{liu2024enhancing} discusses vulnerabilities and cyber-resiliency of DER-based smart grids, \cite{ye2021review} provides a comprehensive overview of attacks on photovoltaic systems and their mitigation methods, and \cite{li2022cybersecurity} surveys existing studies on the cybersecurity of smart inverters for DERs. However, to our knowledge, this work is the first to provide a holistic view and insights into demand-side attacks against power grids, highlighting the cyber vulnerabilities, the impact of the attack, and defense strategies. The key contributions of this work are as follows.

\begin{itemize}
\item We assess the cyber vulnerabilities in IoT-enabled load devices and provide practical examples of their presence in commercially-installed ESAs. More importantly, we relate how these vulnerabilities can be exploited by an attacker to launch LAAs. 

\item By analyzing power grid control loops under LAAs, we highlight the key factors affecting the physical impact of the attack. 

\item We present solutions to enhance the cyber-physical resilience of power grids against IoT-enabled LAAs. We provide an overview of the ongoing regulatory and standardization efforts across different nations to counter demand-side threats. Finally, we present key recommendations to grid operators and policy-making organizations, distilling insights from our research and discussions with industry and policy experts.

\end{itemize}

The remainder of the paper is organized as follows: Section \ref{sec:TARA} presents the threat and risk analysis, while Section \ref{sec:Attack Impact Analysis} focuses on the attack impact analysis. Measures for enhancing cyber-physical system resilience are discussed in Section \ref{sec: Enhancing Cyber-Physical System Resilience}. Section~\ref{sec: Ongoing Regulatory Efforts In Energy-IoT Landscape and Our Recommendations} reviews ongoing regulatory efforts in the energy-IoT landscape and offers our recommendations. Finally, Section \ref{sec: Conclusion} concludes the paper and outlines directions for future research.

\begin{table}[!h]
    \centering
        \caption{Vulnerabilities in commercially-installed ESAs and their exploitation to launch LAAs.}
    \resizebox{0.95\textwidth}{!}{
    \begin{tabular}{|p{4.7cm}|p{4cm}|p{9 cm}|}
    
    \toprule
    \hline
        {\bf Vulnerability} & {\bf Practical Example}  & {\bf Exploitation for LAA} \\
        \hline
       Weak Authentication   &  CVE-2024-4622, CVE-2023-0863  & Allow privilege escalation attacks, enabling direct control by unauthorised entities. Adversaries can then modify load-control settings (e.g., switch on/off air conditioners) leading to large-scale load changing. \\
         \hline
          Remote Code Execution   &  CVE-2024-7795  & Allow malicious entities to execute arbitrary code on remote machines, changing load control functions. \\
          \hline
          Restriction Bypass   &  CVE-2022-22807  & Change load control functions within devices at scale to cause grid instability. \\
          \hline
           Insecure Web Interface  &  CVE-2023-29115, CVE-2023-29120  & Remote load control capability at scale potentially leading to grid instability. \\
          \hline
          Wireless medium &  CVE-2022-0878  & Electromagnetic interference for CCS-based EV chargers interrupting control communication between the vehicle and charger, causing charging sessions to abort. Attacks targeting large EV feet can cause large-scale load changes. \\
          \hline
          Side-channel vulnerabilities &  N/A  & The knowledge of sensitive parameters such as the type and the number of EVs connected, battery capacity and the battery state of charge (derived from side channels) can be used to determine the amount of EV chargers required to be compromised in order to cause a load change of specific magnitude.  \\
          \hline
          \bottomrule
    \end{tabular} 
    }
    \label{tbl:Vulnerabilities}
    \vspace{1cm}
\end{table}

\section{Threat and Risk Analysis}
\label{sec:TARA}
The digitalisation of power grids and their symbiotic integration with Internet of Things (IoT) technologies presents a complex operational landscape replete with various threats. To effectively navigate this realm, a nuanced understanding of these threats and their inherent risks is essential. 

\subsection{Key Vulnerabilities in ESAs and their Exploitation to Launch LAAs}
First, we enlist the cyber vulnerabilities in ESAs. In Table~\ref{tbl:Vulnerabilities}, we provide examples of the vulnerabilities in real-world ESAs (such as EV chargers, air conditioners, etc.), and illustrate how they can be exploited to impact power grid operations.
\begin{itemize}
    \item {\bf Weak Authentication Mechanisms:} A considerable number of IoT devices, such as EV chargers and smart home appliances, are hampered by suboptimal authentication frameworks. 
    \item {\bf Lack of Security Updates:} Many IoT devices are not privy to regular software enhancements, exposing them to augmented risk from cyber incursions.  This is a particular concern for ESAs, which, unlike many other consumer IoT devices, are expected to have a 10-year plus lifetime.
    \item {\bf Remote Code Execution (RCE):} Alarmingly, many IoT devices lack stringent access control measures. Poor control measures mean that users may be able to access an account that should only be able to have limited permissions, but poor control allows that account to do far more than should be enabled.  This can lead to vulnerabilities such as RCE. 
    \item {\bf Insecure Web Interface:} Devices with inadequately secured web interfaces are vulnerable to unauthorized access, presenting a significant risk. This can lead to remote load control capability at scale, potentially leading to grid instability. 
    \item {\bf Vulnerability due to the Wireless Medium:} ESAs such as EV chargers can be vulnerable to attacks using the wireless medium. An example is the `BROKEN-WIRE'' attack, which disrupts charging sessions by exploiting weaknesses in the channel access mechanism \cite{kohler2022brokenwire}. A weak preamble signal transmitted wirelessly in proximity to the chargers deceives the EV and charger modems into believing the communication channel is continuously busy, forcing them to defer data transmission indefinitely.
    \item {\bf Side-Channel Vulnerabilities:} Side-channel attacks involve the exploitation of indirect information to infer sensitive parameters of the system. For instance, the current exchanged during the charging phase on an EV can be used to identify and profiling EVs \cite{brighente2024evscout2}.
\end{itemize}

\begin{figure}[t]
    \centering
    \includegraphics[width=0.8\linewidth]{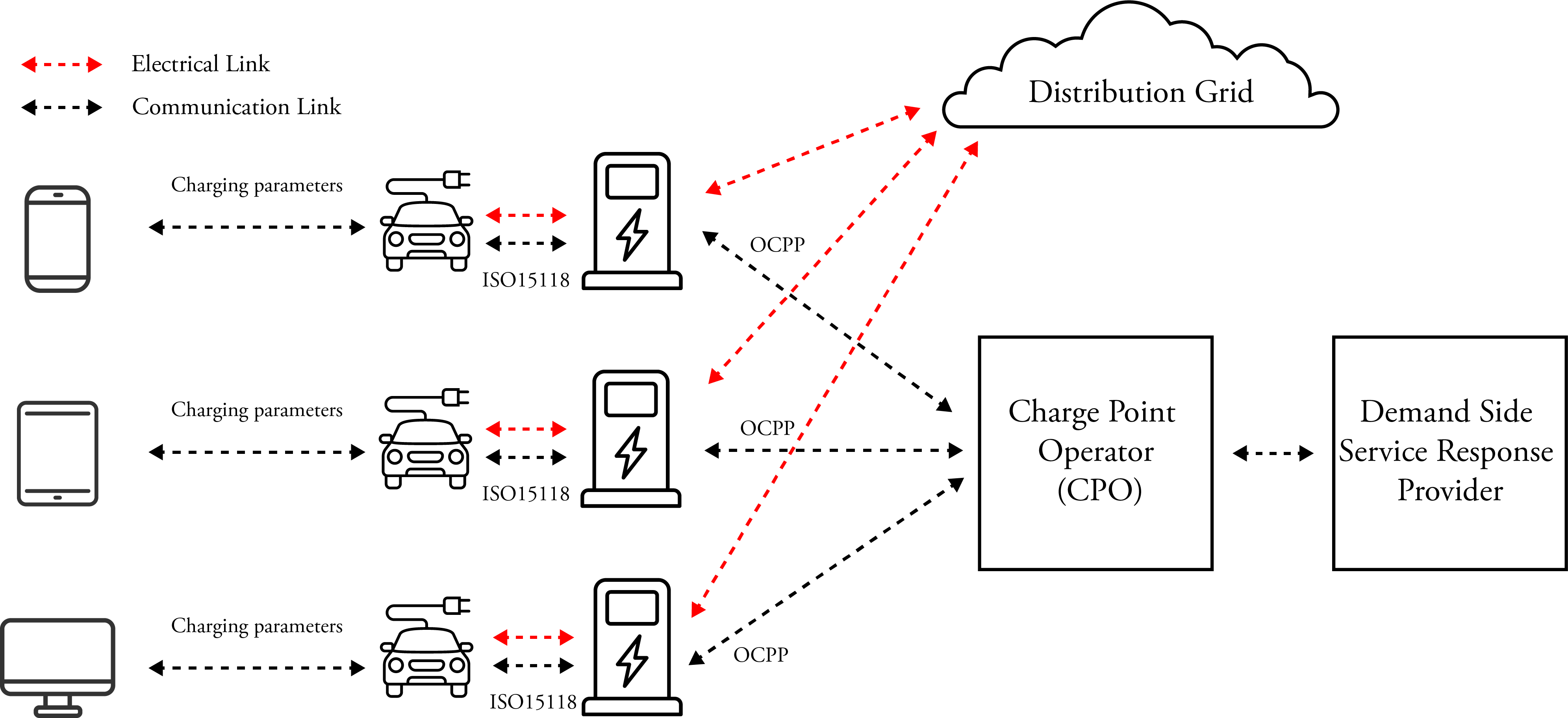}
    \caption{Electric vehicle charging system. ISO-15118 and OCPP refer to the communication protocols used in EV-EVCS and EVCS-central server communication. }
    \label{fig:EVCS_Fig}
    \vspace{1cm}
\end{figure}

While the vulnerabilities listed above are focused specifically on the ESAs, there is a wider risk of access to the load-controlling systems from the communication interfaces involved in DSR operations. Take the example of a smart electric vehicle charging station (EVCS) shown in Fig. ~\ref{fig:EVCS_Fig}.  They can be targeted in several ways (other than compromising the EVCS themselves).
\begin{itemize}
    \item Compromising the communication interfaces between ESAs, CEMs, and the demand-side service providers. E.g., the OCPP protocol used for communication between the EVCS to the CS is known to be vulnerable to man-in-the-middle (MitM) attacks \cite{Alcaraz2017}.
    \item Compromising smart-phone applications used by consumers to control the charging operation.
    \item Social-engineering attacks targeting consumers leading to synchronised customer behaviour.
\end{itemize}

Other than exploiting the vulnerabilities, an attacker may also require auxiliary information to design their LAAs, such as the knowledge of the power grid topology, real-time data on the power demand, electricity price, location information of ESAs, etc.
For example, to calculate the threshold load that causes grid frequency excursions, the attacker will require knowledge of the grid topology, line reactances, etc. These can be obtained by open-source databases \cite{keliris2019open} or blind topology estimation techniques \cite{grotas2019power}. 
Additionally, EV charger smartphone applications, especially those provided by third-party aggregators may reveal the relevant information such as the type of chargers, occupancy, and the power demand \cite{acharya2020public}.

\subsection{Attack Vectors}

The honeypot data analysis also sheds light on potential attack vectors.
\begin{itemize}
    \item {\bf Load-Altering Attacks:} A pivotal threat in the domain of IoT, LAAs have the capacity to modify the load-controlling settings of IoT devices, thereby causing pronounced load fluctuations. Such alterations could have a direct bearing on power grid dynamics. For instance, vulnerabilities that allow escalation of user privileges can enable adversaries to remotely control devices, potentially inducing significant instability in the power grid.
    \item {\bf Denial of Services (DoS \& DDoS):} These attacks are meticulously designed to overwhelm target systems. DDoS attacks, in particular, are more dispersed and pose a greater challenge due to their origin from multiple sources. Such attacks can cause disruption to the systems or devices used in load control leading to loss of DSR services which could impact grid operations.
\end{itemize}
In the next section, we discuss the impact of LAAs on power grids, which portrays the tangible repercussions of these threats on power grid dynamics. 

\section{Attack Impact Analysis} \label{sec:Attack Impact Analysis}
The power grid frequency is an indicator of the balance between the supply (generation) and the demand. Sudden and abrupt manipulation of the power grid demand due to large-scale cyber attacks against ESAs can disrupt this balance and lead to severe effects, such as unsafe frequency excursions, eventually leading to cascading line and generator outages. In the following, we list some key factors influencing the impact of LAAs.

\begin{itemize}
    \item {\bf Magnitude of Vulnerable Load:} For LAAs to be impactful, an adversary must be able to alter a significant amount of load since power grid design philosophies, such as N-1 contingency scheduling, provide resilience against component loss, such as the loss of a generator or a transmission line as well as large-scale load changes. Our studies suggest that LAA magnitudes in the range of hundreds of megawatts can severely impact the power grid's safety, potentially leading to cascading failures \cite{GoodridgeRare2023}. 
    \item {\bf Duration Over which the Load Changes Occurs:} The smaller the duration over which the load change occurs, the greater the attack impact. For instance, the only mitigation against a sub-second load change is the system’s inertial response. On the other hand, if the load change occurs over a relatively longer duration (tens of seconds to a minute), fast-acting local frequency responses such as battery energy storage systems can be leveraged to correct the imbalance (see Section 6). For slower load changes (those that take more than several minutes), the operator can redispatch the generators to compensate for the load changes (i.e., bring new generation sources online). 
    \item {\bf Spatial Factors:} Another key finding is that the impact of an LAA depends critically on the geographical location of the ESAs that are targeted under a cyber-attack \cite{LakshIoT2021}. In other words, an identical magnitude of load change at different locations will impact power grid operations differently. Furthermore, a coordinated attack from multiple locations in the grid can severely threaten the grid’s safety. 
    \item {\bf Temporal factors:} The risk factor due to LAAs varies as a function of the diurnal, weekly or seasonal load conditions. During the peak demand periods (e.g., evening period), the power grid is operating close to its capacity limits, and even a relatively small increase in loads can overwhelm the system and trigger emergency actions. Moreover, strategic attackers can also coordinate their attacks during extreme weather events (such as storms or flood disruptions), during the time when the power grid is already under extreme stress. 

\begin{figure}[t]
    \centering
    \includegraphics[width=0.7\linewidth]{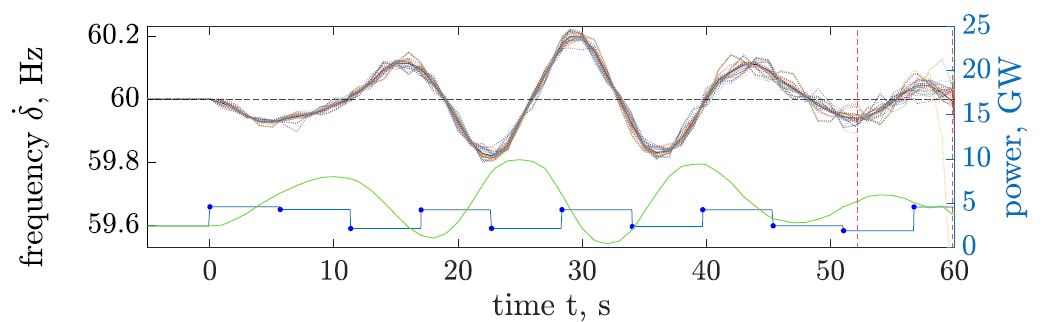}
    \caption{Example of a one-time step attack (top figure) and oscillatory attack (bottom figure). Network frequency profiles (black), network generation profiles (green), and network load profiles (blue). Load changes are depicted with a blue marker and emergency responses as vertical red dashed lines at the time of their occurrence.  (Figure source \cite{GoodridgeRare2023})}
    \label{fig:Osc_attack}
    \vspace{1cm}
\end{figure}

    \item {\bf Oscillatory Load Changes:} While a one-time surge or drop in the demand is undesirable, sophisticated attack patterns such as oscillatory load variations (periodic increase and decrease in demand) can seriously threaten the power grid stability (see Fig. ~\ref{fig:Osc_attack}). Studies have found that such strategic load variations can induce power grid failures by manipulating a lower fraction of the demand as compared to one-time load changes. 
\item { {\bf Renewable Energy Penetration Conditions:} {The growing integration of renewable energy resources will further exacerbate the vulnerability of power grids to LAAs. Conventional generators (such as those based on fossil fuels) have large inertia, which plays a critical role in slowing down the rate-of-change-of-frequency following a disturbance. In contrast, solar and wind energy sources lack the inherent capability to provide inertia to the power system.}

}
\end{itemize}

\section{Enhancing Cyber-Physical System Resilience} \label{sec: Enhancing Cyber-Physical System Resilience}
Given the growing penetration of ESAs in the foreseeable future and the lack of a current unified approach to implementing security standards, there is no silver bullet to eliminating demand-side cyber threats. Although regulators, device manufacturers, and operators are in the process of identifying potential LAA threats and introducing appropriate regulations, there will remain a risk of a successful attack. Thus, operators must prepare strategies to counter the adverse effects of such attacks. This can be done in three stages: by developing protection (preventive), detection, and response and recovery mechanisms. 

\subsection{Protection Measures}

First, we enlist some approaches that can be designed to protect the system and prevent the risk and the impact of LAAs.

\begin{itemize}
\item {\bf Randomization and Ramp rate constraints:} As explained in Section III (second bullet point), the duration over which the load change occurs is an important factor in determining the attack impact. In order to cause a large-scale load change over a short duration, an attacker must be able to alter the load control settings of a large number of devices in a synchronized manner. This effect can be mitigated if a randomized offset is added to the device's start/end times (i.e., a delay whose value is chosen randomly is added between the time a device receives the load control command and the time that the command is actually executed). In the UK, a 10-minute randomization interval is prescribed for EVCSs. 
\item  {\bf Frequency responsive loads:} The negative effects of LAAs can be mitigated quickly if the balance between supply and demand can be restored. A potential way to do accomplish this is to set frequency response functionality in the design of load devices. This could help manage fluctuations with devices modulating their own load if they detect frequency changes. E.g., when a load device senses a drop in the power grid frequency, it can limit the power drawn from the grid and vice versa. 
\item  {\bf Verifying Load Control Commands:} An anomaly detection service could be used to correlate the original flexibility requests sent from the transmission grid operator (e.g., National Grid) to the demand-side service providers, with the requests that they, in turn, send to the end devices. This would enable the detection and (where appropriate) blocking of anomalous load control messages by ensuring that the amount of load control requests sent, correlates broadly with the original instruction. 
\end{itemize}

\vspace{-0.3 cm}

\subsection{Detection Measures}
LAAs can be detected and localized by deploying a cyber-physical monitoring framework either at the bulk power or distribution grid levels.

\subsubsection{Bulk Power Grid Level} At the bulk power grid level, measuring the end-user demand in real-time is not practical. Nevertheless, LAAs can be localized by monitoring the power grid dynamics (i.e., voltage phase angle and frequency fluctuations) and inferring the attack parameters/attack location using machine learning algorithms \cite{JahangirIoTJ2023}. Such a detection mechanism can be seamlessly integrated into existing wide-area monitoring systems, as shown in Figure~\ref{fig:PMU_Hierarchy2}.
A pre-trained CNN an provide inference results extremely quickly. For instance, in the IEEE-57 bus transmission system, the CNN can provide online inference within 89 ms \cite{JahangirIoTJ2023}.
However, for practical implementation, the monitoring framework would require the extensive deployment of phasor measurement units that can provide a sub-second measurement update rate, but their deployment in real power systems is limited.

{
Sharing the data to a central server can also raise cyber security concerns. Thus, it is important to ensure that grid operators comply with existing smart grid communication security standards, such as the IEC 62351 standard, which provides security recommendations for the IEC 61850 protocol used in PMU/substation communication.}

\subsubsection{Distribution Grid Level}
A monitoring framework deployed at the distribution grid level can provide a more fine-grained localisation of LAAs, e.g., identify the compromised ESAs or the distribution network feeders. However, the challenge is that distribution networks are monitored poorly, unlike transmission networks. Moreover, power utilities do not have direct access to all the IoT’s data. Data integrity and customer privacy are significant concerns when it comes to the use of IoT data.

Federated learning can help tackle the data scarcity issue and protect user data privacy by not bringing the raw user data out of the privacy-protected area. In this context, both cyber and physical features from IoT devices can be leveraged, such as source/destination IP, source/destination port, packet length, protocols, intra-packet arrival time, load, temperature setpoint, indoor area, building thermal insulation, power generation, rating, solar irradiance, charging/discharging rate, battery state of charge (SoC), etc. By applying federated learning and incorporating these diverse features, the distribution network operator can calculate an overall IoT trustability score (ITS) while concurrently safeguarding the privacy of IoTs  \cite{Sarker2023}. 

Several important issues require consideration to enable the implementation of the FL-based approach in real-world systems.
\begin{itemize}
    \item {\bf Scalability and Energy Efficiency}  
    can be an important issue with an increasing number of ESAs and data traffic. Moreover, implementing FL over resource-constrained ESAs can be a challenge. Several techniques can be applied to make FL scalable. 
    Asynchronous updates can used in FL to reduce communication overhead, minimize delays, and reduce network load \cite{mcmahan2017}. Optimized scheduling with minimal frequency and careful data selection can reduce energy consumption while avoiding redundant training on outdated data. 
    Besides, personalized FL approaches and robust aggregation methods can help address the challenges posed by uneven data distribution across devices in a large-scale system \cite{li2020federated}. Moreover, FL training can be executed through effective client selection and optimal resource utilization in scale systems with a large number of heterogeneous devices\cite{chen2020joint}. Additionally, the memory requirement for deploying FL in ESAs can be reduced using techniques such as efficient gradient checkpointing\cite{sohoni2019low} and tensor rematerialization\cite{jain2020checkmate}.

    \item {\bf Latency:}  Communication overhead can be a contributor to latency. This can be minimized using asynchronous updates \cite{xie2019asynchronous} and gradient sparsification\cite{lin2017deep}, which reduce the volume of data transmitted during model updates. Additionally, hierarchical FL frameworks can be employed, where local aggregations occur at intermediary nodes before updating the global model \cite{liu2020client}. This reduces the frequency and volume of global communication, significantly improving response times.
\end{itemize}

\vspace{-0.3 cm}

\subsection{Respond and Recover}
Responding to the attack, i.e., attack mitigation, in turn, involves two phases. In the short term (a few seconds to minutes), actions must be taken to prevent the system from destabilising and moving toward cascading failures. At a longer timescale (tens of minutes to hours), the attack must be isolated, and the system must be restored to a safe operational state.

In the short term, power from fast-acting resources (such as battery energy storage systems) can be an effective source to mitigate the imbalance caused by LAAs. Modern-day power grids are already facing significant power imbalances (with or without cyber attacks), for example, instantaneous changes in power balance due to tripping of generation or interconnectors, and minute-by-minute imbalances due to the increasing penetration of renewables. The power grid operators are already introducing several measures to deal with these challenges. (i) The UK's National Grid is in the process of introducing faster-acting frequency response products, namely, dynamic regulation, dynamic balancing, and dynamic containment. The aim of these services is to quickly mitigate imbalances and keep the grid frequency around the setpoint of 50Hz. (ii) Use of zero-carbon synchronous compensators to deliver inertia to the grid. (iii) Use of offshore wind and vehicle-to-grid functionality to provide short-term grid balancing and grid stability services. While these measures can also be applied to stabilize the imbalance created due to LAAs, mitigating cyber attacks will be significantly more challenging, which we discuss in Section VI. 

\subsection{Real-world Testbeds to Simulate LAAs}

\begin{figure}[!ht]
    \centering
    \includegraphics[width=0.9\linewidth]{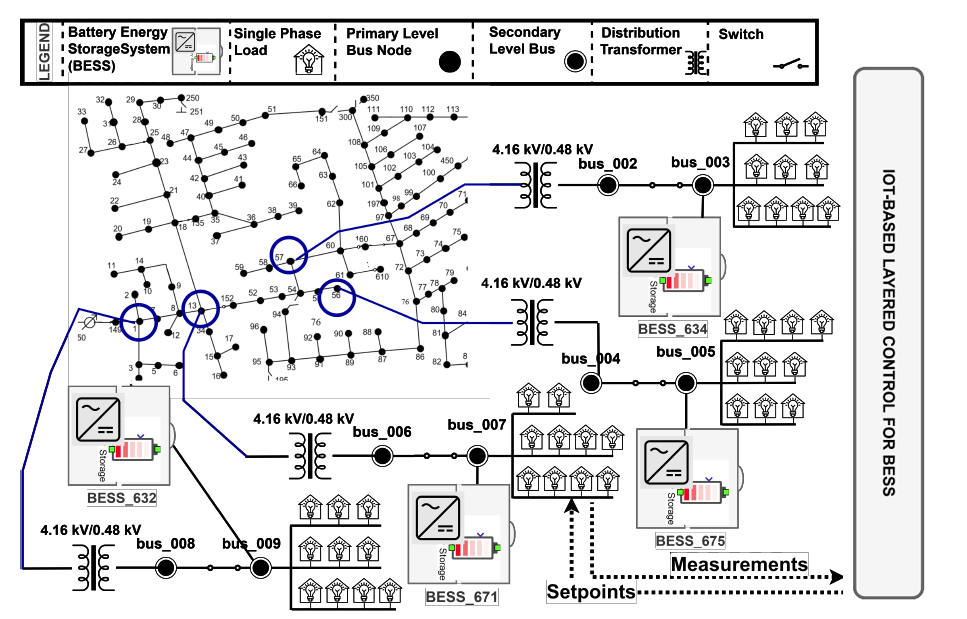}
    \caption{Overview of IoT test bed at West Virginia University}
    \label{fig:IoT_Testbed}
\end{figure}

To understand the impact of the attacks and mitigation measures precisely, it is necessary to develop high-fidelity testbeds that can model the power grid operations with ESAs. To the best of the authors' knowledge, real-world experiments to emulate load-altering attacks (LAAs) are lacking. Nevertheless, in this subsection, we provide the functional requirements for a power grid IoT system testbed based on our setup at the Smart Grid Resiliency and Analytics Lab (SG-REAL) at West Virginia University as shown in Fig.~\ref{fig:IoT_Testbed}, and then highlight some directions to scale up the testbed to simulate LAAs and implement the proposed defence measures. 

For real-world experimentation, we require a co-simulation platform that models the distribution network, IoT devices, associated controllers, and the communication network. We use IoT-enabled battery energy storage systems (BESS) as an example ESA and the IEEE-123 bus distribution system to model the power grid, which is implemented using the OpenDSS platform. The complete BESS model is created using the Typhoon HIL Control Center V2024.1 suite to interconnect with four secondary bus nodes of the IEEE 123-bus distribution network. A central controller collects sensor measurements from individual IoTs. It generates set points (active and reactive power rate) for each of the BESS to ensure support for the connected loads or to charge the battery through the connected primary node of the grid. A two-layer IoT network is built using Mininet to demonstrate the communication among the centralized control center, IoT
devices, and corresponding BESS systems.  MQTT protocol, which is a popular IoT-based lightweight protocol has been used to emulate communication between the IoT-hosts and the respective BESS operating in Typhoon HIL.  

For modelling the cyber attack, the attacker starts performing stealthy network scanning to gather information about the network and devices. After the reconnaissance phase,  with a phishing attack, the attacker gains access to one of the authorized IoT hosts which allows them to intercept, modify, and relay messages between the aggregator and the IoT devices without getting detected. The attacker can exploit the vulnerabilities in the communication network to manipulate the battery reference setpoints, thereby causing incorrect battery charge-discharge decisions. The impact of the attack on the grid is simulated using the OpenDSS platform, resulting in nodal voltage violations. The proposed federated learning-based algorithm can be implemented within this setup as well.

While the proposed testbed emulates an example of an attack targeting small-scale ESAs, simulating LAA requires scaling up the setup to include several thousands of ESAs connected to the grid, which can incur a significant computational burden. Moving forward, some innovative approaches such as multi-fidelity simulations can be useful. An example is to simulate fast-switching ESAs using electromagnetic transient solvers (EMT) while the rest of the system is simulated at a slower time step such as the transient system analysis, thus reducing the overall computational burden \cite{jinsiwale2023design}. 

\begin{figure}
    \centering
    \includegraphics[width=0.7\linewidth]{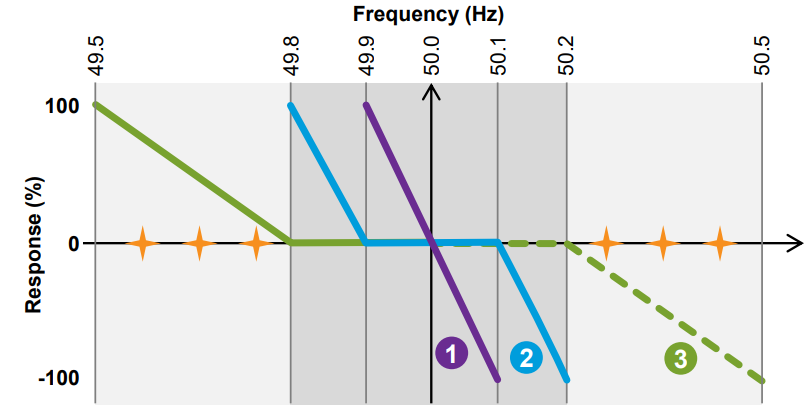}
    \caption{National Grid’s new frequency response services, 1. Dynamic regulation, 2. Dynamic balancing, and 3. Dynamic containment (Source: National Grid ESO) }
    \label{fig:NG_FS}
    \vspace{1cm}
\end{figure}

{
\section{Ongoing Regulatory Efforts In Energy-IoT Landscape and Our Recommendations} \label{sec: Ongoing Regulatory Efforts In Energy-IoT Landscape and Our Recommendations}

With DSR and ESAs becoming key elements in decarbonizing the electricity sector, several nations are in the process of formulating regulations aimed at minimizing associated cyber risks. 

\begin{itemize}

\item The United States has introduced ``Cybersecurity Labeling Program for Smart Devices'' with the objective of helping consumers choose ESAs that are safer and less vulnerable to cyberattacks. The government is committed to enhancing IoT cybersecurity by leveraging Federal research and development (R\&D), procurement, and risk management initiatives, as outlined in the IoT Cybersecurity Improvement Act of 2020. 
\item UK’s Department of Energy Security and Net Zero (DESNZ) proposes the requirement of ESAs to comply with minimum cyber security standards using the ETSI 303 645 framework and bringing large-scale load-controlling organizations (a threshold of $300$ MWs is applied) under National Information and Systems regulations \cite{BEIS_Consult}. 
\item The European Union has approved an  ``action plan for the digitalization of the energy system'', which includes establishing requirements to facilitate data access for demand response, as well as to support the adoption of ESAs.  
 \item The Australian Electric Market Operator 
is taking a holistic view of software management and cyber security risks for distributed energy resources, including the security of communications and controls, establishing nationally aligned ramp rates, and/or randomized delay requirements for ESAs. 
            
\end{itemize}

\vspace{-0.3 cm}
    
    \subsection{Key Recommendations Based on Our Research}
    Next, we present key recommendations to power grid operators and policy-making organizations in this sector, split into recommendations for the cyber layer and the physical layer.  
    
    {\bf Cyber Layer:} Robust cyber security features must be deployed at every level of DSR services. First, there is already a proposition that ESAs are built according to good device standards, such as ETSI 303 645. This standard has requirements for many of the above contained within its 13 principles, including the requirements to minimize exposed attack surfaces. Second, organizations involved in DSR, such as ESA manufacturers and service providers, must ensure that organisational security controls such as ISO 27001/2 are implemented appropriately and that any platforms are also appropriately secured. It is also expected that service providers/ESA manufacturers monitor these ESA device logs and alerts for anomalous activity and potential security incidents and act on them accordingly. Likewise, any material security incidents must be alerted to the consumer in some way, e.g., a new account was created, a password was changed, or an unauthorised change was detected in the software.}

    Interoperability with different power grid platforms and ESA devices is an important factor in fully realising the benefits of DSR. A critical first step toward achieving this is the establishment of defined standards, underpinned by regulatory mandates outlining minimum requirements. Without these standards, organizations will continue to rely on bespoke communication protocols or vendor-specific APIs, which, while functional, are often not designed or secured (in the case of ESAs) for large-scale DSR applications. This creates cybersecurity risks in communication and management systems for ESAs and restricts consumers' ability to switch DSR service providers. Consequently, this results in lock-in barriers for consumers and increases the complexity for service providers, who must integrate with numerous different vendor APIs. To overcome these challenges, it is important to require a common data and information model, along with open communication protocol standards for both power grid platforms and ESA devices. This would eliminate current constraints and enable a secure and functional DSR ecosystem. An example of how this can be achieved is by adopting a minimal data model and specifying standard protocols such as OpenADR and OCPP for communication. Additionally, the UK Government’s Interoperable Demand Side Response (IDSR) program is an ongoing effort to demonstrate the effectiveness of this approach. It consists of three key workstreams to support the innovation, design, and demonstration of interoperable DSR systems, in alignment with the BSI PAS standards (BSI PAS 1878 and 1879). While this program exemplifies how interoperability can be achieved, other technical specifications and protocols are also under consideration by industry and governments to find the right balance between a secure and interoperable DSR system.
    
    {\bf Physical Layer:} Section~III enlists various factors influencing the impact of LAAs. They are particularly relevant in the context of the regulatory efforts described in Section V-A, such as prescribing randomization/ramp rate limits and critical load threshold limits. For example, currently, single static threshold values are recommended for these parameters without regard to spatial/temporal or the oscillatory attack context. Grid operators must perform feasibility studies to carefully determine the parameters of load threshold/randomization/ramp rate limits considering all the factors described above, potentially making the threshold dependent on the location, time, and frequency of load changes.

    Moving forward, an integrated approach will be needed to secure power grids from demand-side threats that include not only preventive measures (such as those listed above) but also attack detection and mitigation features. For attack detection, fusing information from cyber (e.g., network logs, etc.) and physical sources of the power grid physical signals (e.g., voltage, current measurements, etc.) can be a promising approach. However, more research is needed to address the associated practical issues (see Section~VI). For mitigating these threats, energy storage and fast-acting inverter-based resources can be effective sources.

    {\bf Human Factors:} IoT-enabled load devices involve end users (customers) who operate the loads.  Thus, considering human factors is important in assessing and mitigating the threat due to cyber attacks. For instance, when an operator identifies an issue with the large-scale IoT-enabled load devices involving cyber compromise, the user must be immediately alerted to make the necessary changes, such as encouraging them to change the device passwords or apply software updates. End users are untrained and may typically lack security awareness. For example, certain demographics, such as older adults, tend to exhibit low engagement with cyber security prescriptions (such as password updates/two-factor authentication, etc.) due to a combination of low self-efficacy and lack of awareness. Thus, it is important for DSR and grid operators to perform studies on human behavioural factors and adopt security/threat mitigation policies that are compatible with end-user know-how. This must be an important area of future research.
    
\section{Conclusions and Future Research Directions} \label{sec: Conclusion}

In this article, we presented a comprehensive overview of the security vulnerabilities in ESAs, the physical impact of large-scale load changes, and measures to detect and localize these threats. 
We conclude the manuscript by highlighting future research directions.

{
\begin{itemize}
    \item First, existing works on this topic either study the impact of LAAs on the transmission grid (i.e., its impact on the supply-demand balance and grid frequency, while ignoring its impact on the underlying distribution grid) or the localized effects of LAAs on the distribution grid (e.g., potential voltage violations, etc.). To the best of our knowledge, a detailed investigation of the LAA's impact on integrated transmission-distribution systems has not been studied. In particular, understanding how the effect of LAA propagates from distribution to the transmission grid and the associated frequency dynamics requires a detailed investigation.
    \item The use of SCADA and PMU data described in Section III-B can only approximately geolocate the source of an LAA (e.g., identify the substation from which the LAAs have originated), but it does not help identify the specific ESAs involved in the attack. Further research is required to identify the data sets, tools and techniques that could be used for post-LAA investigation and root cause analysis. A potential approach could be requiring devices to send their security logs to a central ``Security Operations Centre'' and correlate that with the bulk grid monitoring. However, this will likely need some consumer consent and privacy notice when signing up for DSR services.
    \item Innovative control techniques need to be developed to adapt existing frequency stabilization methods to mitigate cyber-attacks, since the mitigation responses are tightly coupled with attack detection (e.g., using the monitoring framework described above). Thus, the mitigation must deal with the uncertainties in the detection/localization results (such as false positives and misdetections) as well as detection delays. Furthermore, the interactions between the cyber and the physical layer must be considered during restoration. 
    \item Finally, the long-term attack isolation problem is a challenge that has still not received sufficient attention and much research will be needed in this direction. Looking forward, identifying large loads and having a secure backup communication channel for manual override of ESAs in the event of imminent outages (even if it's devolved to more local areas), may be an option. However, consumer impacts would need to be carefully considered as well as the risk that control itself introduces.
  
\end{itemize}
\vspace{-0.4 cm}
}

\section{Biographies}
{\bf Subhash Lakshminarayana} is an associate professor at the University of Warwick, UK. His research interests include cyber-physical system security, power grid optimization/control and wireless communications. He serves as Associate Editor at the IEEE Internet of Things and IET smart grid journal. His works have been selected for the Best Paper Award for IEEE Smartgridcomm 2024 and Top 5 papers published in IEEE Transactions on Smart Grid 2021-22. His research has been funded by EPSRC, Innovate UK and Horizon Europe.

{\bf Yexiang Chen} received his B.Sc. degrees from Hohai University and the University of Strathclyde in 2017, followed by an M.Sc. from the University of Strathclyde in 2019. He completed his Ph.D. at the University of Warwick in 2024. Currently, he serves as a Postdoctoral Research Fellow in the School of Engineering at the University of Warwick, U.K. His works focus on the developing deep learning techniques to detect cyberattacks and analyze vulnerabilities in electric vehicle charging infrastructure. His research interests include the application of artificial intelligence and moving target defense to support power grid operations and security.

{\bf Carsten Maple}
is the Principal Investigator of the NCSC–EPSRC Academic Centre of Excellence in Cyber Security Research, University of Warwick,
where he is a Professor of Cyber Systems Engineering with Warwick Manufacturing Group. He is also a Fellow of the Alan Turing Institute,
the National Institute for Data Science, and AI in the U.K., where he is a Principal Investigator on a project developing a trustworthy national identity to enable financial inclusion.

{\bf Andrew Larkins}
is founder and CEO of Sygensys. He is working with a wide range of industry stakeholders, including grid operators, regulators, and OEMs to maximise the benefits of mass adoption of smart charging, vehicle to grid and other demand side response systems. 
Previous experience includes developing innovative electronic products and IoT systems across multiple sectors from medical devices to semiconductors.

{\bf Daryl Flack}
is an expert in smart energy cyber security and specialises in consulting for the UK Government on large scale Smart Energy, Critical National Infrastructure (CNI) programmes. He is also the co-founder and CISO of BLOCKPHISH, a leading Managed Security Service Provider (MSSP) with a dedicated Energy Security Practice. Daryl is certified to the highest level by the National Cyber Security Centre (NCSC) - a part of GCHQ. 

{\bf Christopher Few}
is a cyber risk assessor at National Grid and was previously a cyber researcher at Ofgem.  He is a member of the Chartered Institute of Information Security and the IET.
Cyber security has been a theme through his career for over 30 years.

{\bf Kenny-Awuson David} a Principal Cyber Researcher, Data Service and Cyber at Ofgem. He is a distinguished principal cyber researcher at Ofgem Cyber and AI Directorate, E14 4PU London, U.K., bringing
over a decade of expertise in cybersecurity and defense. With nine years of
operational experience in the British
Army, he applies a mission-critical
mindset to advancing cyber resilience. As a Senior Research Fellow at
De Montfort University, he specializes
in cyberphysical systems, leveraging
his Ph.D. in blockchain cloud forensics
from Coventry University. He is also
an academic journal guest editor for
the Remote Sensing Society of Japan.  

{\bf Anurag K. Srivastava} is a Raymond J. Lane Professor and Chairperson of the Computer Science and Electrical Engineering Department at the West Virginia University. He is also an adjunct professor at Washington State University and a senior scientist at the Pacific Northwest National Lab. He is an IEEE Fellow and IEEE Distinguished lecturer, and the author of more than 350 technical publications including a book and 3 patents.  His research interest includes data-driven algorithms for power system operation and control including cyber-resiliency analysis. 

\begingroup
\fontsize{12}{14.4}\selectfont
\bibliographystyle{IEEEtran}
\bibliography{IEEEabrv,bibliography}
\endgroup

\vspace{-1.3 cm}

\end{document}